\documentclass[aps,prx,superscriptaddress,reprint]{revtex4-2}
\usepackage{amsmath}
\usepackage{amssymb}
\usepackage{graphicx}
\usepackage{url}
\usepackage{hyperref}
\usepackage{color,xcolor}
\usepackage{epstopdf}
\usepackage{float}
\usepackage{ulem}
\usepackage[percent]{overpic}
\usepackage{siunitx}

\usepackage{dcolumn}% Align table columns on decimal point
\usepackage{bm}% bold math
\usepackage{times}
\usepackage[version=3]{mhchem} % Formula subscripts using \ce{}

\usepackage{mathrsfs}
\usepackage{lipsum}
\usepackage{amsfonts}
\usepackage{array}
\usepackage{ulem}

\usepackage{lineno}
\usepackage[none]{hyphenat}
\begin{document}

\title{Stability of the novel interorbital-hopping mechanism for ferromagnetism \\ in multi-orbital Hubbard models}
\author{Ling-Fang Lin}
\email{lflin@utk.edu}
\affiliation{Department of Physics and Astronomy, University of Tennessee, Knoxville, Tennessee 37996, USA}
\author{Yang Zhang}
\email{yzhang@utk.edu}
\affiliation{Department of Physics and Astronomy, University of Tennessee, Knoxville, Tennessee 37996, USA}
\author{Gonzalo Alvarez}
\affiliation{Computational Sciences \& Engineering Division and Center for Nanophase Materials Sciences, Oak Ridge National Laboratory, Oak Ridge, TN 37831, USA}
\author{Michael A. McGuire}
%\email{mcguirema@ornl.gov  \\ \\ Notice: This manuscript has been authored by UT-Battelle, LLC under Contract No. DE-AC05-00OR22725 with the U.S. Department of Energy. The United States Government retains and the publisher, by accepting the article for publication, acknowledges that the United States Government retains a non-exclusive, paid-up, irrevocable, world-wide license to publish or reproduce the published form of this manuscript, or allow others to do so, for United States Government purposes. The Department of Energy will provide public access to these results of federally sponsored research in accordance with the DOE Public Access Plan (http://energy.gov/downloads/doe-public-access-plan).}
\author{Andrew F. May}
\affiliation{Materials Science and Technology Division, Oak Ridge National Laboratory, Oak Ridge, Tennessee 37831, USA}
\author{Adriana Moreo}
\author{Elbio Dagotto}
\affiliation{Department of Physics and Astronomy, University of Tennessee, Knoxville, Tennessee 37996, USA}
\affiliation{Materials Science and Technology Division, Oak Ridge National Laboratory, Oak Ridge, Tennessee 37831, USA}

\begin{abstract}
Recently,  it was argued that a ferromagnetic (FM) insulating phase can be induced by a novel {\it interorbital} hopping mechanism. Here, we study the stability range of this novel FM phase under modifications in the crystal fields and electronic correlation strength, constructing a theoretical phase diagram. A plethora of states is unveiled, including the FM Mott insulator (MI),  a FM orbital-selective Mott phase (OSMP), several anferromagnetic (AFM) MI phases, an AFM metallic state, and a FM metal as well. Our most interesting result is that the FM regime, either in MI or OSMP forms, is shown to be stable in {\it large} portions of the phase diagram, at both intermediate and strong electronic correlations, respectively. Our results demonstrate via a detailed example that the recently proposed novel mechanism to stabilize FM insulators is not fragile but instead robust, and may enlarge substantially the relatively small family of known FM insulators.
\end{abstract}

\maketitle

\begin{figure*}[!htb]
\centering
\includegraphics[width=0.96\textwidth]{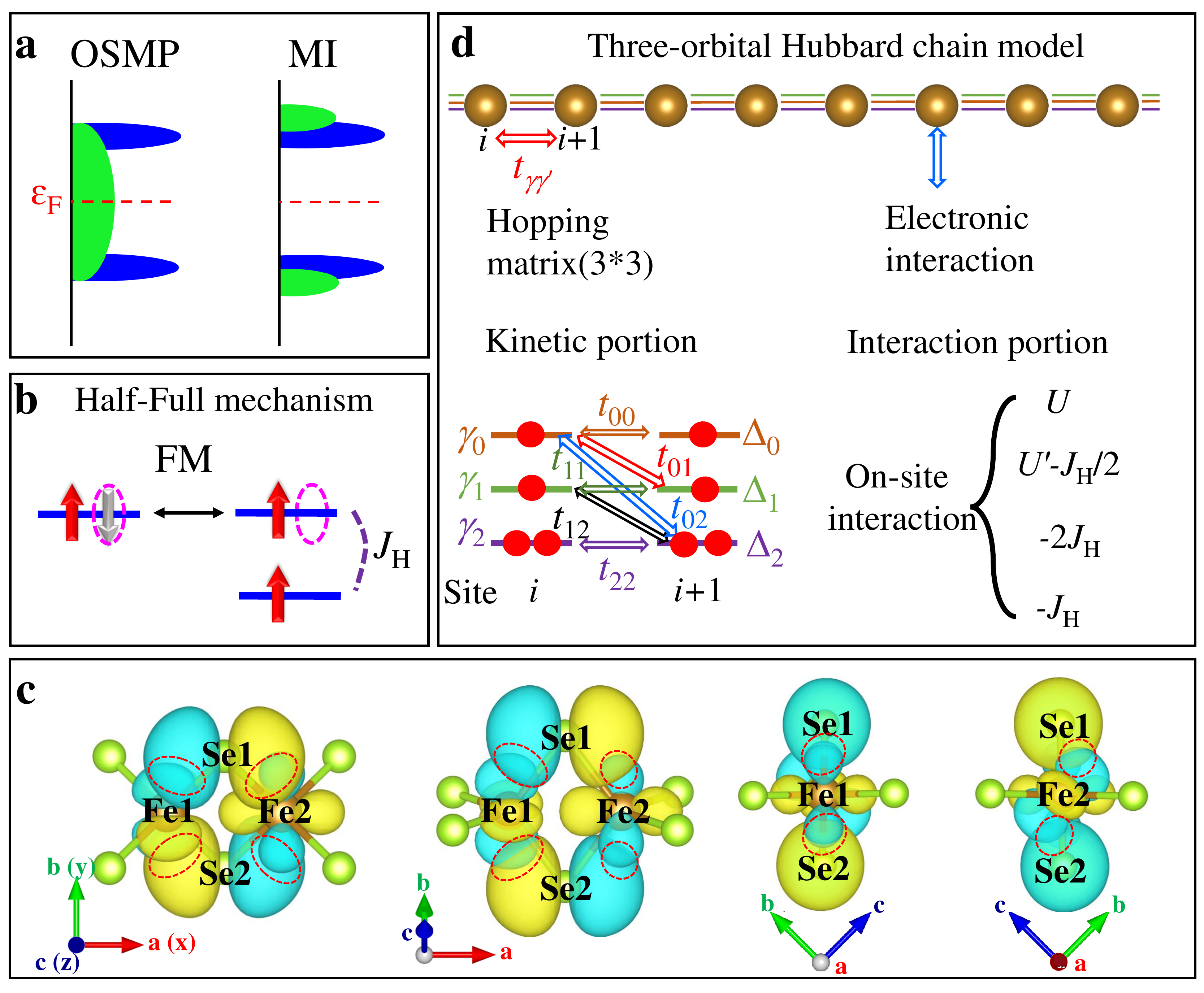}
\caption{{\bf Orbital-selective Mott phase, the novel half-full mechanism and model Hamiltonian.} {\bf a} Sketch of the local density-of-states for the OSMP and MI phases, in a multi-orbital system with electronic correlations. Here we use two electrons in two orbitals per site, as example.
For the OSMP, one orbital forms a Mott state with a gap, while the other orbital remains itinerant and gapless, leading to a globally metallic phase. For the MI state, both orbitals form Mott states, resulting in an insulating phase. {\bf b} Sketch of the FM superexchange mechanism discussed in Ref.\cite{lin2021origin} induced by the large entanglements between doubly occupied and half-filled orbitals. Orbitals are indicated by blue lines. Electrons with spin up or down are indicated by red or grey arrows, respectively. The two-way thin arrows indicate the overlap between inter-site orbitals. The virtual hopping process is highlighted by the magenta dashed ovals. {\bf c} The effective Wannier functions (WF) of orbital $\gamma_1$ ($d_{yz}$) for Fe1 and $\gamma_2$ ($d_{x^2-y^2}$) for Fe2 with bridges using the Se1 and Se2 $p_{x}$ orbitals. The robust overlap between these WFs, which is related to the amplitude of hoppings $t_{12}$, are indicated by the dashed red ovals. Isosurface is set to be 3 here. Different colors represent different signs of the WF. The Wannier function plot is produced using WANNIER90 code~\cite{mostofi2008wannier90} and VESTA~\cite{momma2011vesta}. {\bf d} The three-orbital Hubbard model on a one-dimensional chain lattice geometry used in our study (see text for details).}
\label{Model}
\end{figure*}

\noindent {\bf \\Introduction\\}
Transition metal (TM) systems involving multi-orbital correlated electrons continue attracting much attention due to their rich physical properties~\cite{Dagotto:Rmp94,Scalapino:rmp,Dai:Np,Dagotto:Rmp,Khomskii:cr,ZhangJ0state}.
In the standard Hubbard Hamiltonian of a multiorbital system, the interplay of the elements of the hopping matrix, the crystal fields $\Delta$ splitting orbitals, the Hubbard repulsion $U$, and the Hund coupling $J_H$ linking all orbitals, often leads to several intriguing electronic phases arising from their competition, such as the molecular-orbital state in dimers~\cite{Khomskii:cr}, the spin-singlet state~\cite{Streltsovt:prb14,zhang2021peierls}, various forms of orbital ordering~\cite{Tokura:Sci,lin2021orbital}, and the recently much-addressed orbital-selective physics~\cite{caron2012orbital,zhang2021orbital,yin2011kinetic,kostin2018imaging,zhang2021magnetic,di2022orbital}. An interesting example of orbital-selective states is the unusual orbital-selective Mott phase (OSMP) [see Fig.~\ref{Model}{\bf a}], involving a mixture of localized and itinerant behavior of the different orbitals when in the {\it intermediate} electronic correlation region~\cite{de2014selective,Mourigal:prl,Craco:prb20,LindopingOSMP,StepanovOSMP}.

It is precisely the intermediate coupling regime that harbors the most surprises because in this region hand-waving lines of reasoning are often not trustworthy since many couplings are similar in magnitude and reliable predictions are difficult, unless involving robust computational methodologies. This is the true regime of ``complexity'' in correlated electrons, where the expression complexity is used as denoting the emergence of exotic unexpected properties from seemingly simple local interactions.
This intermediate region will be the focus of our present effort.

Consider now the primary specific
goal of this publication. Most TM insulating materials are antiferromagnetic (AFM) while few display ferromagnetic (FM) order. This can be understood
from the simplicity and robustness of Anderson's superexchange AFM theory which is based on second-order perturbation theory in the hopping amplitudes~\cite{anderson1959new,goodenough1963magnetism}. However, recently a ``half-full'' mechanism involving large entanglements between doubly-occupied and half-filled orbitals was proposed~\cite{lin2021origin} to understand the puzzling origin of FM order along the chain direction in \ce{Ce2O2FeSe2}~\cite{mccabe2011new,mccabe2014magnetism}, as shown in Fig.~\ref{Model}{\bf b}. Note that we employ the word ``entanglement'' between orbitals with different electronic population as indicating that in a quantum description these orbitals cannot be considered as independent of one another, but they are intrinsically coupled
and this interplay of two orbitals with different electronic populations is at the heart of the novel mechanism proposed.

Specifically, based on second-order perturbation theory in the hopping amplitudes, the total gain in energy of the FM configuration
due to the robust hopping $t_{12}$ between half-occupied and fully-occupied orbitals was found to be:
\begin{equation}
\begin{split}
\Delta E_{\rm FM} &=-\frac{|t_{12}|^2}{U-3J_{\rm H}+\Delta}.
\end{split}
\end{equation}
where $\Delta$ = ${\Delta}_1-{\Delta}_2$ is the crystal-field splitting between the half-filled and double-occupied orbitals.
On the other hand, in the AFM state the total energy gained from $t_{12}$ is
\begin{equation}
\begin{split}
\Delta E_{\rm AFM}&=-\frac{|t_{12}|^2}{U-J_{\rm H}+\Delta}.
\end{split}
\end{equation}

Hence, the interorbital electronic hopping $t_{12}$ by itself favors a FM interaction, basically driven by the Hund coupling.
However, the prevailing intra-orbital hoppings $t_{11}$ or $t_{22}$ favor a superexchange AFM state with gains in energy such
as $\Delta E$ $\sim$ -$\frac{|t_{11}|^2}{U+J_{\rm H}}$.
Then, the dominant magnetic order of the material under consideration is decided by the competition of these different channels.
Our previous work showed that in some cases, the hopping $t_{12}$ can be large enough to stabilize a FM insulating phase,
without the need to resort to the more frequently mentioned mechanism of double exchange, which produces robust FM but with metallic character,
and the nearly $90^{\circ}$ bonds, typically associated with low critical temperatures. It is also worth remarking that while
our previous theoretical work, as well as the present effort, are in one dimension, the simple perturbative foundations of these ideas
are robust and valid in higher dimensions as well.

To intuitively understand the effective interorbital hopping $t_{12}$ between the Fe1 $d_{yz}$ and Fe2 $d_{x^2-y^2}$ orbitals, it is crucial
to include Se's $p$ orbitals in the Wannier90 calculations~\cite{mostofi2008wannier90}. By comparing all channels contributing to $t_{12}$,
we found out that the dominant channel occurs when the Se's $p_x$ orbital acts as a ``bridge'' between the $3d$ orbitals.
To have an intuitive visual view of $t_{12}$, the WFs related to this $p_x$ channel are shown in Fig.~\ref{Model}{\bf c}. In this sketch, the bending of the Fe-Se-Fe bond is important to achieve a nonzero $t_{12}$ matrix element.

Considering the previous successful application of this novel idea to a particular material~\cite{lin2021origin}, using specific numbers for hoppings and crystal fields,
several questions naturally arise.
How does the FM state induced by this half-full mechanism evolve by varying the crystal-field values as well as the strength of the electronic correlations?
At intermediate couplings, could the OSMP with FM order be instead stable in this half-full system with crystal field effects?
What other interesting magnetic or electronic phases can emerge by considering the competition of those parameters in an extended phase diagram? In other words,
are the previous results~\cite{lin2021origin} an anomaly or truly indicative of a general novel mechanism?

To address these broad goals, we investigated the crystal field and electronic correlations effects on the new half-full FM ground state previously reported,
by using the density matrix renormalization group (DMRG) technique on the multi-orbital Hubbard model. Fixing the Hund coupling to $J_{\rm H}/U = 0.25$,
realistic for materials of the iron family \cite{Dagotto:Rmp}, the magnetic and electronic phase diagram was theoretically constructed varying $\Delta$ and $U/W$.
One of our main results is that the FM phase is stable in a large portion of the phase diagram. Namely, the previously reported FM phase due to
a robust interhopping amplitude $t_{12}$ is here shown not to be fragile, as often spin liquid states tend to be,
but representative of broad tendencies that until now were not considered
by the community of experts. Specifically, we found both the recently discovered FM Mott insulating (MI) and the FM OSMP phases are stable at intermediate and strong electronic correlation, respectively. Our results potentially open a vast
avenue of research and the possibility for the family of FM insulators to be considerably enlarged. We propose that via {\it ab-initio} techniques,
a systematic exploration of materials with robust interorbital hopping could provide the first steps toward additional FM insulators in the near future.

In addition, several other interesting magnetic electronic phases are
also revealed in our study arising from the competition of hopping, crystal field splitting, and electronic correlations,
involving an AFM2 metal (i.e. an AFM state with blocks of size 2), a FM metallic state, and staggered canonical AFM1 MI phases.
This exemplifies the remarkable complexity that emerges from multiorbital models
when they are analyzed with reliable computational techniques, particularly in the challenging intermediate coupling regime.
%Was AFM2 and AFM1 defined before?

\noindent {\small \bf \\Why FM insulators are important for applications?\\}
%A column of text or less summarizing ferromagnetic insulators and semiconductors for "Stability of the recently proposed ferromagnetic phase induced by interorbital hopping in multi-orbital Hubbard models". We use spintronics applications as a motivation for interest in these materials.
Before moving into the technical aspects, we wish to briefly remind the reader of the practical importance of enlarging the family of FM insulators.
Ferromagnetic (or ferrimagnetic) insulators have important applications in the field of spintronics, where their low magnon damping and ability to exchange couple magnetism into neighboring materials are particularly useful \cite{vzutic2004spintronics, dieny2020opportunities, ahn20202d}. Compared to AFM insulators, FM insulators are relatively rare, particularly among oxides \cite{choi2017highly, frantti2019quest, tsurkan2021complexity}. Ferrimagnetic ferrite garnets like \ce{Y3Fe5O12} (yttrium iron garnet, or YIG) are commonly used in devices; however, some limitations of these materials have been noted and development of alternatives is an active and important area of research \cite{beyondgarnets}. Perhaps the earliest recognized example of a ferromagnetic insulator was \ce{CrBr3} \cite{tsubokawa1960magnetic}. The combination of ferromagnetic order and cleavability has made this compound, its analogue \ce{CrI3}, and the related compounds \ce{CrSiTe3} and \ce{CrGeTe3} important materials in the field of functional van der Waals heterostructures \cite{Zhong2017, lohmann2019probing, ahn20202d}. Although ferromagnetism is usually associated with transition metals, several rare-earth-based ferromagnetic insulators are also known, including the rock-salt structure compounds EuO, EuS, and mononitrides of several rare earth elements \cite{schmehl2007epitaxial, wei2016strong}. Hexagonal ferrites and spinels such as \ce{CoFe2O4} \cite{isasa2014spin, amamou2018magnetic} are also of interest \cite{beyondgarnets}. Ferromagnetic semiconducting behavior has also been reported in spinel chalcogenides like \ce{CdCr2S4} and \ce{CrCr2Se4} \cite{lehmann1966electrical, tsurkan2021complexity}.

\noindent {\bf \\Results\\}
\noindent {\small \bf \\Model Hamiltonian\\}
In this effort, as an example of our broad ideas, we employ a canonical three-orbital Hubbard model defined on a one-dimensional (1D) chain lattice,
including the kinetic energy and interaction terms written as $H = H_k + H_{int}$. The tight-binding kinetic component is
\begin{eqnarray}
H_k = \sum_{\substack{i\sigma\\\vec{\alpha}\gamma\gamma'}}t_{\gamma\gamma'}^{\vec{\alpha}}
(c^{\dagger}_{i\sigma\gamma}c^{\phantom\dagger}_{i+\vec{\alpha}\sigma\gamma'}+H.c.)+ \sum_{i\gamma\sigma} \Delta_{\gamma} n_{i\gamma\sigma},
\end{eqnarray}
where the first term represents the hopping of an electron from orbital $\gamma$ at site $i$ to orbital $\gamma'$ at the nearest neighbor
(NN) site $i+\vec{\alpha}$. $c^{\dagger}_{i\sigma\gamma}$($c^{\phantom\dagger}_{i\sigma\gamma}$) is the standard creation (annihilation) operator, $\gamma$ and $\gamma'$ represent the different orbitals, and $\sigma$ is the $z$-axis spin projection. $\Delta_{\gamma}$ represent the crystal-field splitting of each orbital $\gamma$.

The electronic interaction portion of the Hamiltonian includes the standard intraorbital Hubbard repulsion $U$, the electronic repulsion  $U'$ between electrons
at different orbitals, the Hund's coupling $J_H$, and the on-site inter-orbital electron-pair hopping terms. Formally, it is given by:
\begin{eqnarray}
H_{int}= U\sum_{i\gamma}n_{i \uparrow\gamma} n_{i \downarrow\gamma} +(U'-\frac{J_H}{2})\sum_{\substack{i\\\gamma < \gamma'}} n_{i \gamma} n_{i\gamma'} \nonumber \\
-2J_H  \sum_{\substack{i\\\gamma < \gamma'}} {{\bf S}_{i,\gamma}}\cdot{{\bf S}_{i,\gamma'}}+J_H  \sum_{\substack{i\\\gamma < \gamma'}} (P^{\dagger}_{i\gamma} P_{i\gamma'}+H.c.),
\end{eqnarray}
where the standard relation $U'=U-2J_H$ is assumed and $P_{i\gamma}$=$c_{i \downarrow \gamma} c_{i \uparrow \gamma}$.

Specifically, here we consider a three-orbital Hubbard model with four electrons per site [see the sketch in Fig.~\ref{Model}{\bf d}], where the crystal-field splitting and hopping matrix are adopted from the real iron chain \ce{Ce2O2FeSe2} system as a concrete example~\cite{lin2021origin}. To study the crystal-field splitting effects between half-filled and fully occupied orbitals we fixed the values of $\Delta_0$ and $\Delta_1$ as in \cite{lin2021origin} but vary $\Delta_2$, as well as $U/W$.
In the ``DMRG'' portion of the methods section, the reader can find the specific 3$\times$3 hopping matrix and values of $\Delta_0$ and $\Delta_1$ employed.

\begin{figure*}
\centering
\includegraphics[width=0.96\textwidth]{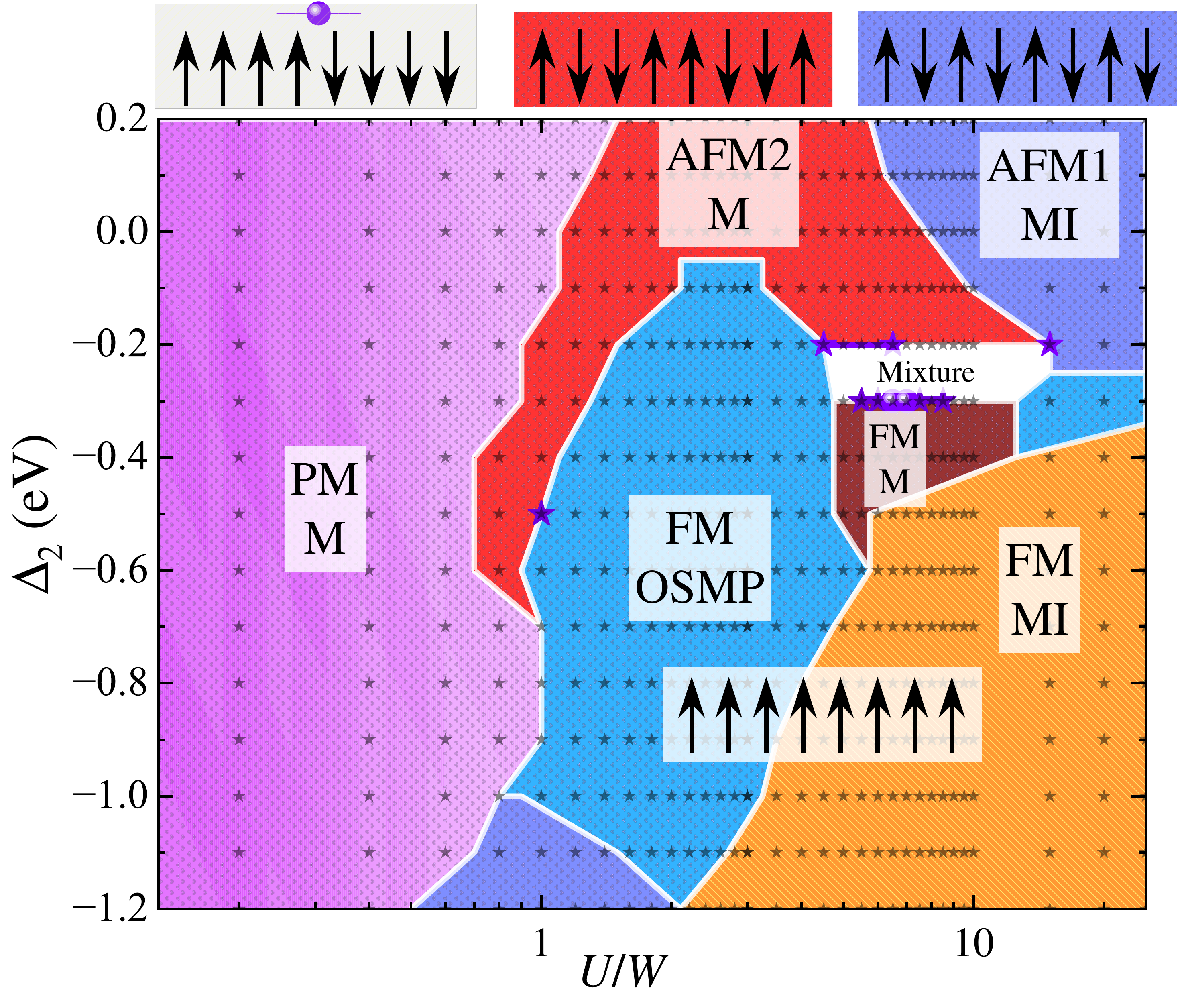}
\caption{{\bf DMRG phase diagram.} Phase diagram of the three-orbital Hubbard model varying $U/W$ and crystal-field splitting $\Delta_{2}$, using DMRG and an $L=16$ chain system with open
boundary conditions. Note that the three-orbital character of the problem renders this system equivalent to a challenging highly entangled 48-site single orbital model.
We use the prototypical value $J_H/U = 1/4$. Different electronic and magnetic phases are indicated by solid regions and labels,
including paramagnetic (PM) metal (M, in pink), FM OSMP (in light blue), AFM2 M (in red), FM MI (in yellow), and AFM1 MI (in purple).
Due to strong competition between opposite tendencies, some mixed phases, marked by purple stars, are found near the boundaries,
especially around $\Delta_{2} \sim -0.25$ eV and $U/W \sim 8$ (white region). The AFM4 phase (blocks of 4), marked by a purple circle, is also displayed in this region.
Note that the boundaries should be considered only as crude approximations. However, the existence of the six regions shown was clearly established by robust DMRG
evidence near their centers, even if the boundaries are only crude estimations. All calculated data points are marked by grey stars.}
\label{phase_cryfield}
\end{figure*}

\noindent {\small \bf \\Phase diagram with crystal field effects\\}
First, using DMRG we constructed the general phase diagram varying the electronic correlations $U/W$ and crystal field splitting $\Delta_{2}$, as shown in Fig.~\ref{phase_cryfield}. This phase diagram is obtained from measurements of the spin-spin correlation $S(r)$, the site-average occupancy $n_{\gamma}$, and charge fluctuations $\delta{n_{\gamma}}$, all at $J_H/U = 0.25$ as already explained. The results are rich and include six different phases: (1) paramagnetic (PM) metallic (M) phase, (2) AFM2 M state,
(3) FM OSMP, (4) AFM1 MI phase, (5) FM M state, and (6) FM MI phase.

At small electronic correlation ($U/W \lesssim 0.7$), the system displays PM behavior with three itinerant orbitals, where the spin correlation $S(r)$ involving two spin operators separated by a distance $r$ decays rapidly with
distance (not shown explicitly for simplicity). In this region, the hopping term plays the leading role, leading to the metallic behavior. At intermediate Hubbard coupling strength, the FM OSMP state -- involving coexisting localized and itinerant electrons -- is found to be stable in a large range of $\Delta_2$. Furthermore, two interesting magnetic and electronic phases are also obtained in this intermediate region by changing $\Delta_{2}$. They are the AFM2 M and AFM1 MI phases, arising from the competition of $\Delta_{2}$ and $U/W$. At large $U/W$, Mott insulating states with localized charges ($n$ number of electrons either 1 or 2) dominate over a broad range of the crystal-field splitting $\Delta_{2}$. Also an FM M phase is found in a small region due to the competition of $\Delta_2$, $U/W$, and hopping amplitudes.

In this rich DMRG phase diagram, the FM phase is dominant in a large region of $\Delta_2$ with the conduction type determined by the strength of the electronic correlations.
At intermediate $U/W$, the system has simultaneously metallic and insulating orbitals, leading to the interesting FM OSMP state.
This OSMP is induced by the competition between hopping values of different orbitals and electronic correlations (Hubbard $U$, Hund coupling $J_H$):
the electrons in the orbital with smaller hopping are {\it localized} at intermediate correlations $U/W$ while the electrons in orbitals with larger hoppings
remain {\it itinerant} i.e. metallic with non-integer filling. If we further increase the electronic correlations, the FM OSMP metallic state
transitions to the FM MI phase where all three orbitals are fully Mott-localized at strong $U/W$. By decreasing the crystal-field splitting $\Delta_2$ towards zero, the FM state changes to different AFM phases because in this regime the canonical AFM superexchange mechanism dominates. Note that there is a small OSMP region at large $U/W$ and $\Delta_2 \sim -0.3$ eV, where orbital $\gamma = 1$ is the localized one. This is slightly different from the large OSMP region at intermediate correlations, where orbital $\gamma = 0$ is the localized one. The reason is that, in the smaller OSMP region, $\Delta_2$ is close to the crystal splitting value of orbital $\gamma = 0$, rendering stronger competition between orbital $\gamma = 0$ and $\gamma = 2$, while $\gamma = 1$ is easier to be localized.

\noindent {\small \bf \\FM-OSMP vs Crystal field splitting\\}
At intermediate Hubbard coupling strengths, the FM OSMP is found to be stable in a large region when decreasing $\Delta_{2}$ into negative values (Fig.~\ref{phase_cryfield}). Let us now focus on the crystal-field splitting effects at intermediate electronic correlation $U/W = 1.6$ in the phase diagram.

\begin{figure}
\centering
\includegraphics[width=0.48\textwidth]{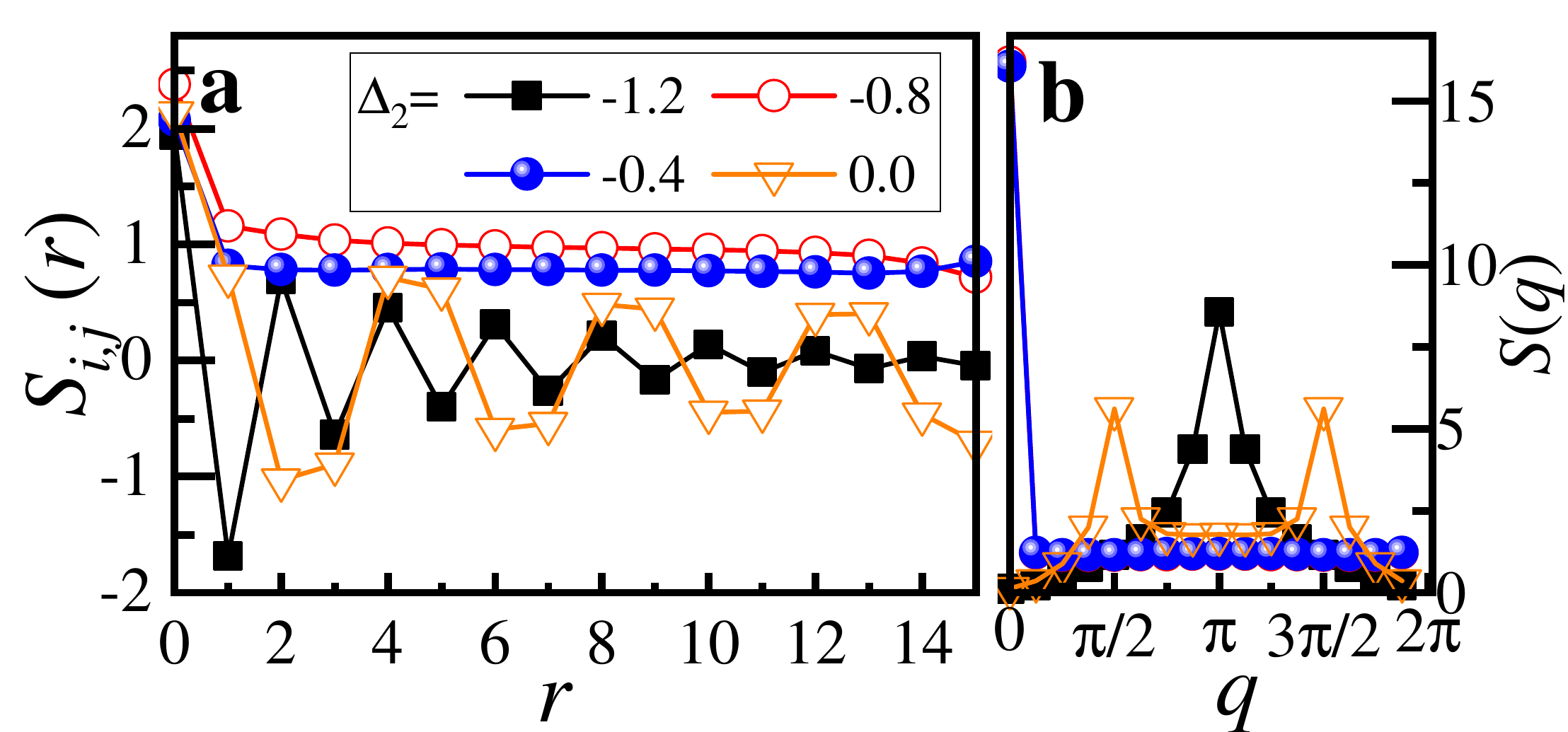}
\includegraphics[width=0.48\textwidth]{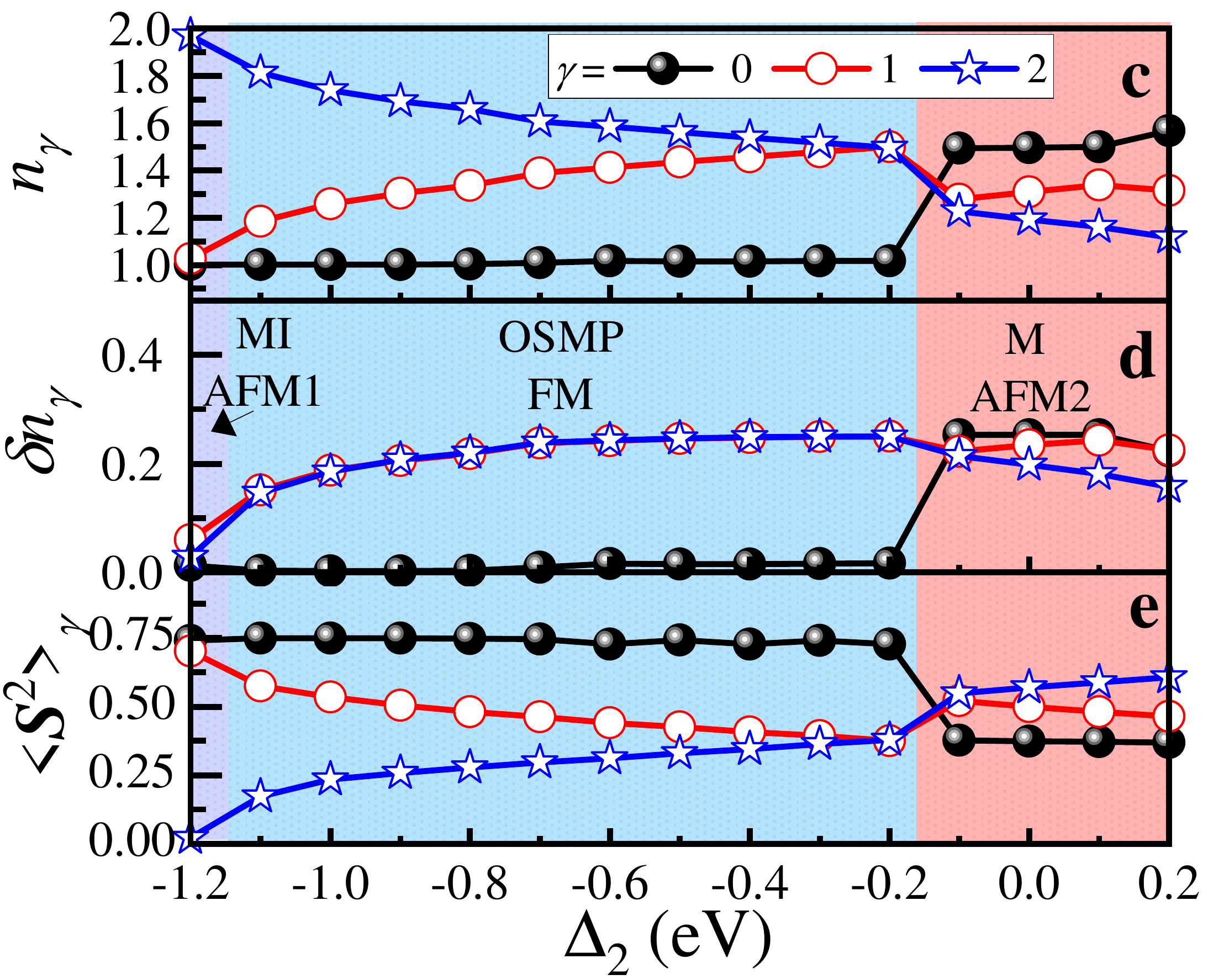}
\caption{{\bf Observables at intermediate correlation.} {\bf a} Spin-spin correlation $S(r)=\langle{{\bf S}_i \cdot {\bf S}_j}\rangle$ (with $r=\left|{i-j}\right|$ in real space) and {\bf b} the spin structure factor $S(q)$, at different values of $\Delta_2$, all at $J_H$/$U$ = 0.25 and $U/W = 1.6$. {\bf c} Orbital-resolved occupation number $n_{\gamma}$. {\bf d} Charge fluctuations $\delta{n_{\gamma}}=\frac{1}{L}\sum_{i}({\langle}n_{i,\gamma}^2\rangle-{\langle}n_{i,\gamma}{\rangle}^2)$. {\bf e} Averaged value of the total spin-squared $\langle{{S}}^2\rangle_{\gamma}$ vs. $\Delta_2$, at $J_H$/$U$ = 0.25 and $U/W = 1.6$. For all these results, the chain length is $L = 16$ and DMRG was used.}
\label{UbyW1.6}
\end{figure}

In the range $-1.1$ eV $\lesssim$ $\Delta_{2}$ $\lesssim -0.2$ eV and at $U/W = 1.6$, the spin-spin correlation $S(r)$=$\langle{{\bf S}_i \cdot {\bf S}_j}\rangle$ vs $r$ indicates
FM order along the chain geometry (see the results at $\Delta_{2}$ = -0.8 and -0.4 eV in Fig.~\ref{UbyW1.6}{\bf a}).
The distance is defined as $r=\left|{i-j}\right|$ with $i$ and $j$ site indexes. As shown in Fig.~\ref{UbyW1.6}{\bf b},
a sharp peak is also observed at $q = 0$ in the spin structure factor $S(q)$, clearly indicating FM order. In addition, the occupation of
orbital $\gamma = 0$ is locked at the integer $1$ in this region of $\Delta_{2}$, leading to the Mott-localized characteristic in this orbital,
while the $\gamma = 1$ and $\gamma = 2$ orbitals have non-integer electronic densities, indicating metallic behavior, as displayed in Fig.~\ref{UbyW1.6}{\bf c}. To better understand the characteristics of the metallic vs. insulating behavior in different orbitals, we also studied the charge fluctuations $\delta{n_{\gamma}}$ for different orbitals, as displayed in Fig.~\ref{UbyW1.6}{\bf d}. In the region ($-1.1$ eV $\lesssim$ $\Delta_{2}$ $\lesssim -0.2$ eV), the $\gamma = 1$ and $\gamma = 2$ orbitals have some charge fluctuations because of the itinerant nature of their electrons. However, the charge fluctuation of $\gamma = 0$ is basically zero due to its localized orbital characteristics. Furthermore, $\langle{{S}}^2\rangle_0$ saturates at $3/4$, corresponding to a half-filled orbital with spin 1/2 at each site,
while $\langle{{S}}^2\rangle_1$ and $\langle{{S}}^2\rangle_2$ are less than 3/4, as shown in Fig.~\ref{UbyW1.6}{\bf e}. Thus, the system displays the orbital selective Mott characteristics with one localized orbital ($\gamma = 0$) and two itinerant orbitals ($\gamma = 1$ and $\gamma = 2$) and with global FM order.

Decreasing further $\Delta_{2}$ ($\lesssim -1.1$ eV), the system displays the canonical staggered AFM phase with the $\uparrow$-$\downarrow$-$\uparrow$-$\downarrow$ configuration unveiled by $S(r)$, while the spin structure factor $S(q)$ shows also a sharp peak at $q = \pi$ (see results for $\Delta_2 = -1.2$ eV in Fig.~\ref{UbyW1.6}{\bf b}).
Moreover, all three orbitals are integer occupied ($n_0 = 1$, $n_ 1 = 1$, and $n_2 = 2$) in this state.
Furthermore, $\langle{{S}}^2\rangle_0$ and $\langle{{S}}^2\rangle_1$ saturates at $3/4$, corresponding to the half-filled orbital, while $\langle{{S}}^2\rangle_2$ is zero, indicating a double-occupied orbital, as shown in Fig.~\ref{UbyW1.6}{\bf e}. Hence, the system is in an AFM1 MI state in this region of $\Delta_{2}$. Because $\Delta = \Delta_{1} - \Delta_{2}$ is larger than the Hund coupling $J_H$, the system can be effectively regarded as a two-half-occupied orbital system, where the inter orbital hoppings lead to the Heisenberg AFM coupling, resulting in the expected MI phase characteristic of one-orbital Hubbard $U$ models, where the Anderson superexchange prevails.

When $\Delta_{2}$ $\textgreater -0.2$ eV, the spin-spin correlation $S(r)$ at $\Delta_2 = 0.0$ eV shows clearly the formation of antiferromagnetically coupled FM spin clusters with a $\uparrow$-$\uparrow$-$\downarrow$-$\downarrow$ AFM2 pattern, as shown in Fig.~\ref{UbyW1.6}{\bf a}. Furthermore, the spin structure factor $S(q)$ also displays
a sharp peak at $q = \pi/2$, corresponding to the AFM2 phase, as shown in Fig.~\ref{UbyW1.6}{\bf b}. Moreover, the four electrons per site
are in non-integer orbitals with large charge fluctuations, indicating metallic behavior [see Figs.~\ref{UbyW1.6}{\bf c} and {\bf d}]. Moreover, the operator
$\langle{{S}}^2\rangle_{\gamma}$ for different orbitals are all less than $3/4$, in agreement with the anticipated
metallic nature of this state, as shown in Fig.~\ref{UbyW1.6}{\bf e}.

\noindent {\small \bf \\The projected density of states (PDOS) $\rho_{\gamma}$ at $U/W = 1.6$\\}
To better understand the many different phases at intermediate electronic correlation, we calculated the orbital-resolved projected density of states $\rho_{\gamma}(\omega)$ vs frequency $\omega$ by using the dynamical DMRG, where the dynamical correlation vectors were obtained using the Krylov-space approach~\cite{Kuhner:prb,Nocera:pre}. A broadening parameter $\eta = 0.1$ was chosen in our DMRG calculations, as in previous cases. The chemical potential is obtained from $\mu = (E_{N+1} - E_{N-1})/2$, where $E_N$ is the ground state energy of the $N$-particle system. $\rho_{\gamma}(\omega)$ is calculated from the portions of the spectra below and above $\mu$, respectively, using:

\begin{equation}
\rho_{\gamma}(q, \omega \textless \mu)=\frac{1}{\pi} Im\left[ \left< \psi_0 \left| c_{i,\gamma}^{\dagger} \frac{1}{\omega +H -E_g
+i\eta}c_{i,\gamma} \right|\psi_0 \right> \right],
\end{equation}

\begin{equation}
\rho_{\gamma}(q, \omega \textgreater \mu)=\frac{-1}{\pi} Im\left[ \left< \psi_0 \left| c_{i,\gamma} \frac{1}{\omega -H +E_g+i\eta}c_{i,\gamma}^{\dagger} \right|\psi_0 \right> \right],
\end{equation}

\begin{equation}
\rho_{\gamma}(\omega)=\rho_{\gamma}(q, \omega \textless \mu)+\rho_{\gamma}(q, \omega \textgreater \mu),
\end{equation}

\noindent with $\psi_0$ as the ground state.

\begin{figure}
\centering
\includegraphics[width=0.48\textwidth]{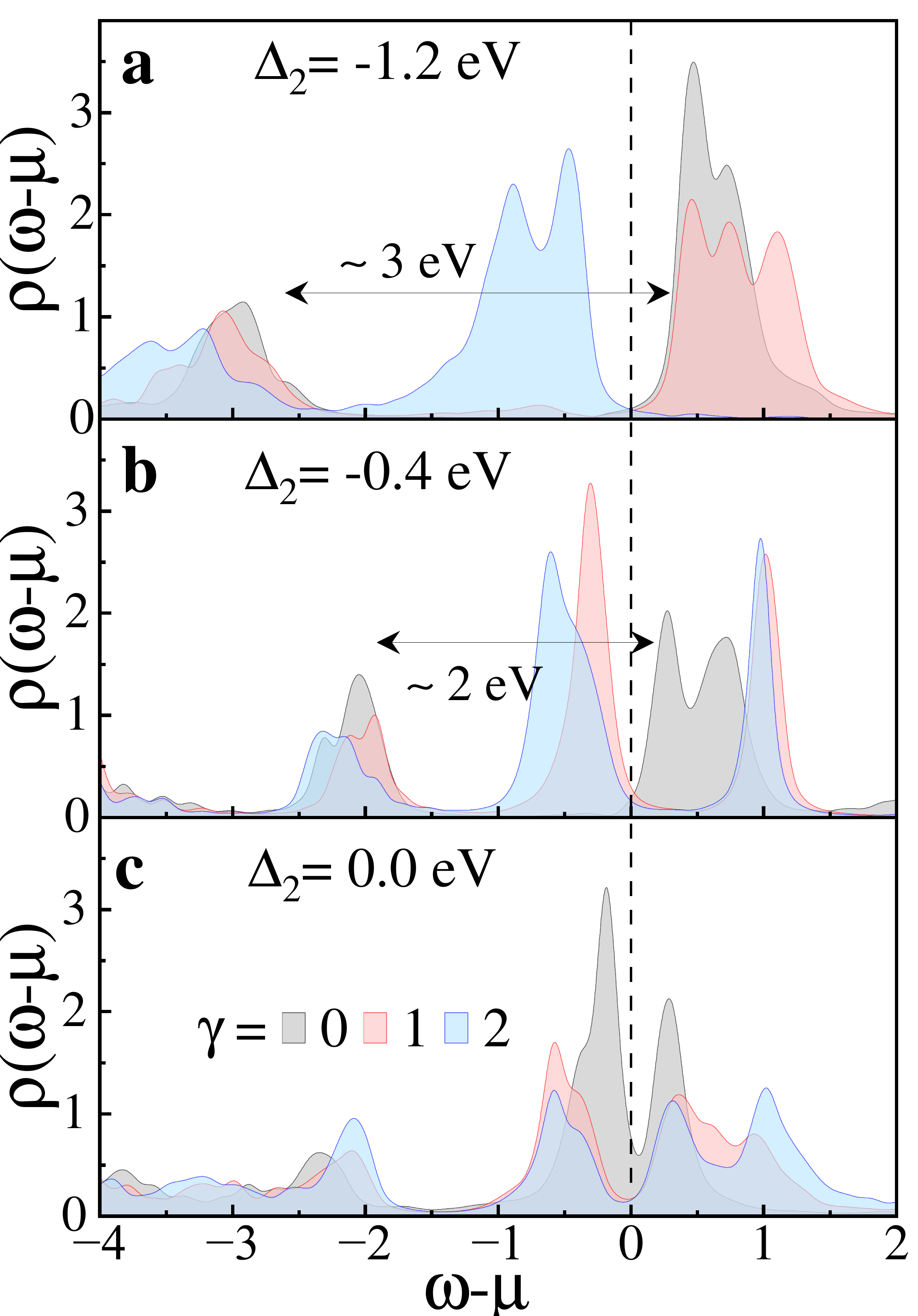}
\caption{{\bf Orbital-resolved density of states.} Density of states (DOS) $\rho_{\gamma}$($\omega$) of different orbitals for different values of $\Delta_2$, at $U/W = 1.6$ and $J_H/U = 0.25$. {\bf a} AFM1 MI state at $\Delta_2$ = -1.2 eV. {\bf b} FM OSMP phase at $\Delta_2$ = -0.4 eV. {\bf c} AFM2 M state at $\Delta_2$ = 0.0 eV.}
\label{dos}
\end{figure}

As displayed in Fig.~\ref{dos}, we calculated $\rho_{\gamma}(\omega)$ for several $\Delta_2$'s, corresponding to different phases, at $J_{H}/U = 0.25$ and $U/W = 1.6$. At $\Delta_2 = -1.2$ eV, the ${\gamma} = 0$ and ${\gamma} = 1$ orbitals display a localized behavior with a Mott gap, while ${\gamma} = 2$ shows fully-occupied characteristics below the chemical potential $\mu$, Fig.~\ref{dos}{\bf a}. Hence, this is a MI state, in agreement with previous discussions based on the site-average occupancy
and charge fluctuations for this region.

At $\Delta_2 = -0.4$ eV, the ${\gamma} = 0$ orbital displays localized behavior with a Mott gap. Meanwhile, the ${\gamma} = 1$ and ${\gamma} = 2$ orbitals have electronic states crossing the Fermi level, leading to metallic behavior, see Fig.~\ref{dos}{\bf b}. Hence, the coexistence of localized and itinerant carriers supports the OSMP picture. By considering the spin-spin correlations $S(r)$ and spin structure factor $S(q)$, this state is a FM OSMP. In fact, the hole-doped manganites are also a FM OSMP system, with $t_{2g}$ orbitals localized and $e_g$ metallic~\cite{SenLa0.7Mn0.3O3}. In doped manganites, FM is induced by the double exchange mechanism~\cite{Dagotto:Prp}, where off-diagonal hopping between the half and empty $e_g$ orbitals play an important role, similar to the half-full hopping discussed here.

Figure~\ref{dos}{\bf c} clearly shows that all three orbitals are metallic with some electronic bands crossing $\mu$, indicating itinerant electronic behavior, leading to an orbitally metallic state. This AFM2 metal was also predicted in the iron selenide chain based on the same model, where the crystal-field splitting $\Delta$ between half-filled and fully occupied orbitals is also small~\cite{pandey2020prediction}.

\noindent {\small \bf \\Strong $U/W$ region\\}
By increasing further the electronic correlation strength, the OSMP itinerant orbitals would become fully Mott-localized by
the Hubbard strength $U$, resulting in a MI phase. To better understand the crystal-field effects on the FM MI state,
we will focus on the discussion of the main results for different values of $\Delta_2$ at strong electronic correlation $U/W = 8$
in the phase diagram.

\begin{figure}
\centering
\includegraphics[width=0.48\textwidth]{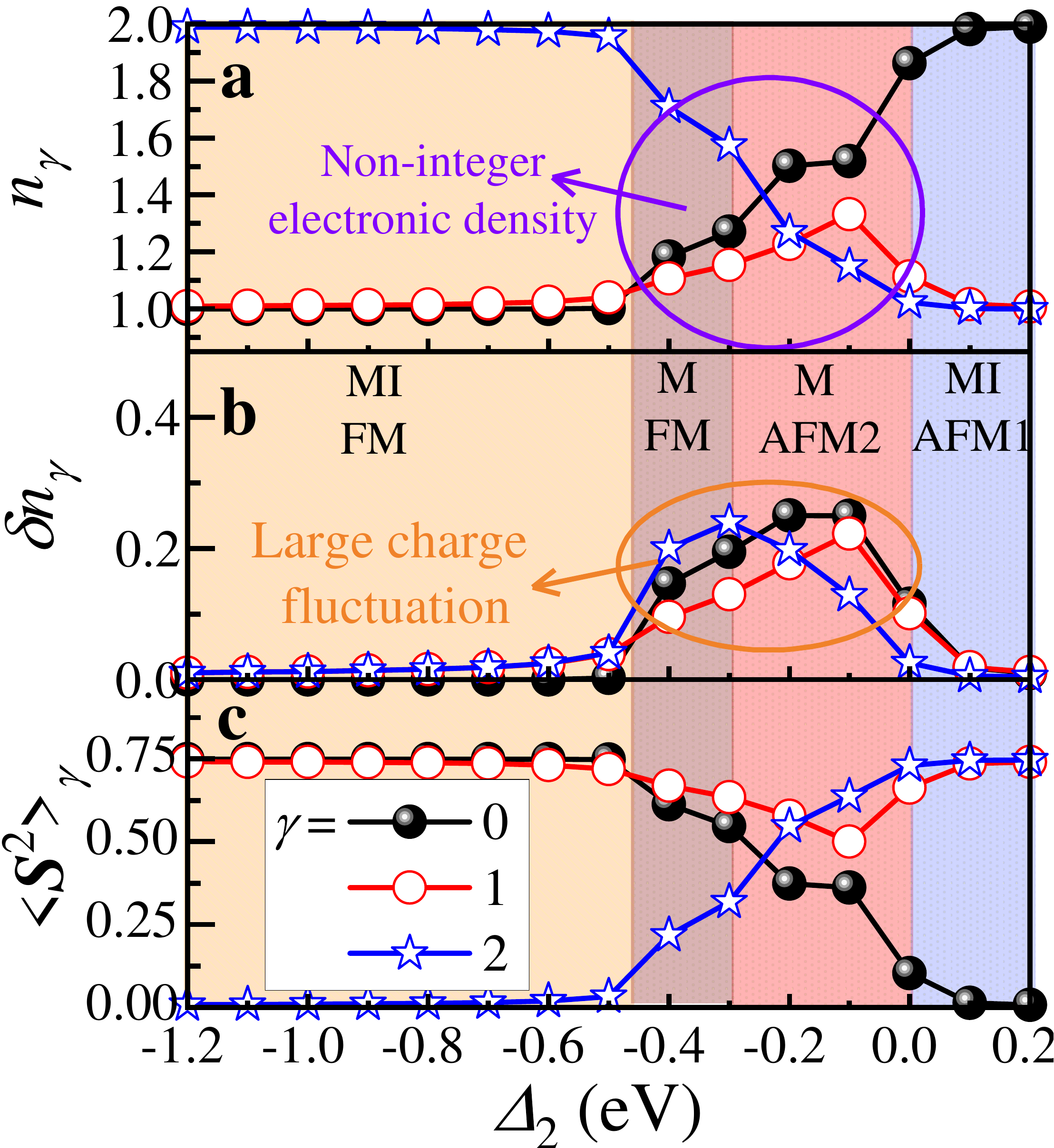}
\caption{{\bf Observables at strong correlation.} {\bf a} Orbital-resolved occupation number $n_{\gamma}$, {\bf b} charge fluctuations $\delta{n_{\gamma}}=\frac{1}{L}\sum_{i}({\langle}n_{i,\gamma}^2\rangle-{\langle}n_{i,\gamma}{\rangle}^2)$, and {\bf c} averaged value of the total spin-squared $\langle{{S}}^2\rangle$, as a function of $\Delta_2$, at $J_H$/$U$ = 0.25 and $U/W = 8$. Here, the length of the chain is $L = 16$ and DMRG was used.}
\label{UbyW8}
\end{figure}

The FM MI state is found to dominate in a large region of $\Delta_2$, as shown in Fig.~\ref{UbyW8}. When $\Delta_2$ $\lesssim -0.5$ eV, the
${\gamma} = 0$ and ${\gamma} = 1$ are half-filled ($n_0 = 1$ and $n_1 = 1$) and ${\gamma} = 2$ is a fully occupied orbital ($n_2 = 2$) without
any charge fluctuation, as shown in Figs.~\ref{UbyW8}{\bf a} and {\bf b}. Meanwhile, $\langle{{S}}^2\rangle_0$ and $\langle{{S}}^2\rangle_1$ are fixed at $3/4$,
while $\langle{{S}}^2\rangle_2$ is zero, indicating a strong Mott-localized behavior, as shown in Fig.~\ref{UbyW8}{\bf c}. Furthermore, the critical Hubbard $U$
for the metal-insulator transition of the FM phase decreases as the crystal-field splitting $\Delta$ = $\Delta_1 - \Delta_2$
(between doubly occupied and half-filled orbitals) increases. In this case, the FM MI state becomes more stable with larger crystal-field splitting $\Delta$.

By increasing the crystal field $\Delta_2$, the system has non-integer electronic density $n_{\gamma}$ in the three orbitals, see Fig.~\ref{UbyW8}{\bf a},
while the charge fluctuations $\delta{n_{\gamma}}$ are large, leading to a metallic phase. In particular, when $\Delta_2$ is in the region,
closer to $\Delta_0 = -0.277$ eV and $\Delta_1 = -0.203$ eV, the competition between orbitals is the strongest.  When $\Delta_2 \gtrsim 0.0$ eV,
the three orbitals begin to localize with integer electronic density $n_{\gamma}$ without charge fluctuations, leading to a MI state, as shown in
Figs.~\ref{UbyW8}{\bf b} and {\bf c}. In addition, the $\gamma = 0$ orbital becomes doubly occupied ($n_0 = 2$), while the  $\gamma = 1$ and $\gamma = 2$ orbitals are half-filled ($n_1 = 1$ and $n_2 = 1$).

\noindent {\bf \\Discussion\\}
In multiorbital systems with more than half-filled orbitals, it was recently shown that the {\it interorbital} hopping between half-filled and fully-occupied orbitals
can lead to an FM insulating phase~\cite{lin2021origin}. To understand how this novel FM mechanism is affected by the crystal-field splitting $\Delta = \Delta_1 - \Delta_2$ between half-filled and fully-occupied orbitals, as well as by the strength of the Hubbard repulsion $U/W$, here we comprehensively studied the $\Delta_2$ vs $U/W$ phase diagram of a three-orbital lattice model defined on a chain, by using DMRG many-body techniques. By modifying the value of $\Delta_2$ over a broad range, the FM phase was found to be quite stable in the phase diagram,
showing that the results of our previous study are not fragile, but indicative of robust tendencies, which also apply in higher dimensions.
The FM state was found to display both FM OSMP or FM MI behavior, at intermediate or strong electronic correlations, respectively.
In addition, several interesting additional magnetic electronic phases were also revealed in our phase diagram arising from the competition of hoppings, crystal fields,
and electronic correlations, involving AFM2 M, FM M, and AFM1 MI states.

Our results indicate that the FM order induced by this half-filled mechanism should be robust. This FM mechanism may explain the FM order in some other materials besides the one we studied before, such as in iron chains with $d^6$ configuration~\cite{eibschutz1975magnetism,toda2005field,stuble2018na7}, one-dimensional $S = 1$ Ni-based chains~\cite{kjems1978evidence}, and Fe$_3$GeTe$_2$~\cite{deng2018gate}, where their unique lattice geometry
provides the possibility of strong overlap between half-filled and fully occupied orbitals, thus creating a robust value for the interorbital coupling. On the other hand, it must be kept in mind that the inter-orbital hoppings ($t_{11}$ and $t_{22}$) lead to AFM tendencies, thus a competition of tendencies will produce the final outcome.
What other interesting phases can be obtained by increasing the values of those inter-orbital hoppings in the large intra-orbital hopping case,
as well as for different electronic correlations remains to be investigated (especially the evolution of the FM OSMP state).

Our study not only focuses on insulators but also on OSMP metals. Because in our case the hopping $t_{11}$ (between $\gamma = 1$ orbitals) is much smaller than others, the $\gamma = 1$ orbitals could easily be localized while the other two orbitals still remain metallic in the intermediate correlated region, leading to the interesting FM OSMP. Recently, a FM OSMP was reported in the Fe$_3$GeTe$_2$ system by neutron experiments~\cite{Fe3GeTe2OSMP-FM}. A more detailed study for the Fe$_3$GeTe$_2$ system would be interesting to perform. Furthermore, in iron-based superconductors, the OSMP is believed by some groups to play a key role to understand superconductivity~\cite{yin2011kinetic,Yi:prl13,yu2013orbital}. Hence, the next step is naturally to try to find additional real materials with the FM OSMP state to study whether superconductivity can be found in that state, complementing our search for FM insulators based on the new mechanism here discussed.

All these future directions of research require systematic work based on density functional theory (to calculate hopping and crystal fields and to identify the relevant orbitals) plus model calculations for correlations. Our study indicates the half-full mechanism could lead to robust FM order under effects of crystal-field splitting. Furthermore, AFM2 M and AFM1 MI phases were also found in our DMRG study at intermediate or strong Hubbard strengths, where those phases were also found or predicted in some other iron chain systems with the same electronic density. Our work is a natural starting point point for a variety of future studies to realize, both in theory and experiment, the important role of intra-hopping between half-filled and full occupied orbitals. This
area of research has been barely touched until now.

In order to find new strong FM insulating states experimentalists should focus on materials with large orbital entanglement and strong crystal-field splitting between half-filled and fully occupied orbitals. The interesting FM OSMP also can be obtained in a range of appropriate crystal field splittings, as in our calculations. Furthermore, crystal field splitting sensitively depends on chemical bonds and crystal structures, which could be in practice tuned by strain or pressure~\cite{liang2022high,byrne2012piezochromism}.

\noindent {\bf \\Methods\\}
\noindent {\small \bf DMRG method\\}
The model Hamiltonian discussed here was studied by using the density-matrix renormalization-group (DMRG) method~\cite{white1992density,schollwock2005density}, where the DMRG++ computer package was employed~\cite{alvarez2009density}. In our DMRG calculations, we used an $L = 16$ sites cluster chain geometry with open-boundary conditions (OBC). In addition, the electronic filling $n = 4$ in the three orbitals was considered. Furthermore, at least $1200$ states were kept during our DMRG calculations and up to $21$ sweeps were performed during the finite-size algorithm evolution. Truncation error remains below $10^{-6}$ for all of our results. An example input file and additional details could be found in the supplemental material \cite{supp}.

\noindent {\small \bf \\The kinetic part of the Hamiltonian\\}
The hopping matrix for the three-orbital chain system is defined~\cite{lin2021origin} in orbital space as follows:

\begin{equation}
\begin{split}
t_{\gamma\gamma'} =
\begin{bmatrix}
          0.187	    &  -0.054	   &       0.020	   	       \\
          0.054	    &   0.351	   &      -0.349	   	       \\
          0.020	    &   0.349	   &      -0.433	
\end{bmatrix}.\\
\end{split}
\end{equation}
The crystal field splitting of the two orbitals are fixed as $\Delta_{0} = -0.277$ and $\Delta_{1} = -0.203$ eV,
while $\Delta_{2}$ was allowed to vary in our DMRG calculations. The total kinetic energy bandwidth $W$ is 2.085 eV.
All parameters mentioned above, involving the hopping matrix and crystal fields, were extracted from our previous work~\cite{lin2021origin}.

\noindent {\small \bf \\Observables\\}
To obtain the phase diagram of the three-orbital 1D Hubbard model varying $U/W$ and $\Delta_2$, several observables were measured using
the DMRG many-body technique.

The spin-spin correlation is defined as:
\begin{eqnarray}
S_{i,j}=\langle {\bf{S}}_{i}\cdot {\bf{S}}_{j}\rangle.
\end{eqnarray}
where ${\bf{S}}_{i} =\sum_{\substack{\gamma}}{\bf{S}}_{i,\gamma}$.

The corresponding structure factor for spin is:
\begin{eqnarray}
S(q)=\frac{1}{L}\sum_{\substack{j,k}}e^{-iq(j-k)}\langle {\bf{S}}_{k}\cdot {\bf{S}}_{j}\rangle,
\end{eqnarray}

The site-average occupancy of orbitals is:
\begin{eqnarray}
n_{\gamma}=\frac{1}{L}{\langle}n_{i\gamma\sigma}\rangle.
\end{eqnarray}

The orbital-resolved charge fluctuation is defined as:
\begin{eqnarray}
\delta{n_{\gamma}}=\frac{1}{L}\sum_{i}({\langle}n_{\gamma,i}^2\rangle-{\langle}n_{\gamma,i}{\rangle}^2).
\end{eqnarray}

The mean value of the squared spin for each orbital is defined as:
\begin{eqnarray}
\langle {\bf{S}}^2\rangle_{\gamma} =\frac{1}{L}\sum_{\substack{i}} \langle {\bf{S}}_{i,\gamma}\cdot{\bf{S}}_{i,\gamma}\rangle.
\end{eqnarray}

\noindent {\bf {\small \\ Data availability\\}} The data that support the findings of this study are available from the corresponding author upon request.

\noindent {\bf {\small \\ Code availability\\}} The computer codes used in this study are available at \href{https://g1257.github.io/dmrgPlusPlus/}{https://g1257.github.io/dmrgPlusPlus/.}

\noindent {\bf {\small \\Acknowledgements\\}}
The work of L.-F.L., Y.Z.,  M.A.M., A.F.M., A.M., and E.D. was supported by the U.S. Department of Energy (DOE), Office of Science, Basic Energy Sciences (BES), Materials Sciences and Engineering Division. G.A. was partially supported by the scientific Discovery through Advanced Computing (SciDAC) program funded by U.S. DOE, Office of Science, Advanced Scientific Computing Research and BES, Division of Materials Sciences and Engineering.

\noindent {\bf {\small \\ Author contributions\\}} L.L., Y.Z. and E.D. designed the project. L.L. carried out numerical calculations for the multiorbital Hubbard model. G.A. developed the DMRG++ computer program. L.L., Y.Z., M.A.M., A.F.M., A.M., and E.D. wrote the manuscript. All co-authors provided useful comments and discussion on the paper.

\noindent {\bf {\small \\ Competing interests\\}} The authors declare no competing interest.

\noindent {\bf {\small \\ Additional information\\}}
Correspondence should be addressed to Ling-Fang Lin or Yang Zhang.

\end{document}